\documentclass[prl,aps,twocolumn,superscriptaddress, 10pt]{revtex4-2}

\usepackage{amsmath}
\usepackage{amssymb}
\usepackage{graphicx}
\usepackage{threeparttable,booktabs}
\usepackage{yhmath}
\usepackage{enumerate}

\usepackage[utf8]{inputenc}
\usepackage{bm}
\usepackage{appendix}
\usepackage{ulem}

\usepackage{bbm}
\usepackage{xcolor}

\graphicspath{{plots/}}

\newcommand{\R}{{\mathbb R}}

\newcommand\1{\leavevmode\hbox{\rm \small1\kern-0.35em\normalsize1}}

\begin{document}
\title{Neural Renormalization Group Flow for Percolation}
\author{A. Alvez}
\affiliation{LISN, AO team, B\^at 660 Universit\'e Paris-Saclay, Orsay Cedex 91405}
\author{L. Camagna}
\affiliation{LISN, AO team, B\^at 660 Universit\'e Paris-Saclay, Orsay Cedex 91405}
\author{S. Chibbaro}
\affiliation{LISN, AO team, B\^at 660 Universit\'e Paris-Saclay, Orsay Cedex 91405}
\affiliation{Inria Saclay - Tau team, B\^at 660 Universit\'e Paris-Saclay, Orsay Cedex 91405}
\author{C. Furtlehner}
\affiliation{Inria Saclay - Tau team, B\^at 660 Universit\'e Paris-Saclay, Orsay Cedex 91405}
\affiliation{LISN, AO team, B\^at 660 Universit\'e Paris-Saclay, Orsay Cedex 91405}
\author{F.P. Landes}
\affiliation{LISN, AO team, B\^at 660 Universit\'e Paris-Saclay, Orsay Cedex 91405}
\affiliation{Inria Saclay - Tau team, B\^at 660 Universit\'e Paris-Saclay, Orsay Cedex 91405}
\author{G. Manzan}
\affiliation{Inria Saclay - Tau team, B\^at 660 Universit\'e Paris-Saclay, Orsay Cedex 91405}
\affiliation{LISN, AO team, B\^at 660 Universit\'e Paris-Saclay, Orsay Cedex 91405}
\author{L. Mensi}
\affiliation{LISN, AO team, B\^at 660 Universit\'e Paris-Saclay, Orsay Cedex 91405}

\begin{abstract}
Machine learning offers a possible route to data-driven real-space renormalization when the relevant observables are nonlocal and difficult to prescribe explicitly. We explore this idea for two-dimensional site percolation developping a supervised, scale-shared neural architecture. The model recursively applies the same learned coarse-graining rule across scales, producing a latent field from which the crossing probability is predicted, while a corresponding fine-graining decoder reconstructs the largest-cluster mask. Trained only on small lattices, the model extrapolates to substantially larger systems, recovers the spanning cluster with high fidelity, and produces observables obeying the expected finite-size scaling near the critical point. We observe that to get such performance it is key that the learned latent representation exhibits critical fluctuations and scale-dependent flows consistent with the renormalization-group structure of percolation. 
\end{abstract}


\maketitle
\paragraph*{Introduction}
The use of machine learning (ML) in physics has led to important progress across many domains~\cite{carleo2019machine}. 
Yet physical datasets differ from generic image or text datasets in a crucial respect: their relevant structures are often constrained by symmetries, conservation laws, causal relations, and long-range correlations. These constraints are not merely useful prior information; in many systems they determine which observables are physically meaningful. This motivates hybrid approaches in which neural architectures are endowed with appropriate physical information, rather than relying exclusively on generic data-driven models~\cite{bronstein2021geometric}.

Scale-invariant systems provide a particularly stringent test case. Processes such as avalanches~\cite{ramos2011scale}, earthquakes~\cite{olami1992self}, and solar flares~\cite{hamon2002continuously} exhibit power-law statistics and correlations across a broad range of scales. Learning such systems requires more than detecting local patterns: the model must represent how observables transform under coarse graining, possibly with nontrivial anomalous dimensions. This requirement is closely related to known limitations of standard neural architectures, including spectral bias~\cite{rahaman2019spectral}, and has recently been emphasized in the context of scale-invariant regression and extrapolation~\cite{alvez2026learning}. In such problems, successful extrapolation is expected only when the architecture or representation captures the underlying self-similarity~\cite{sosnovik2019scale}.

\begin{figure*}[t]
\centering
\includegraphics[width=\textwidth]{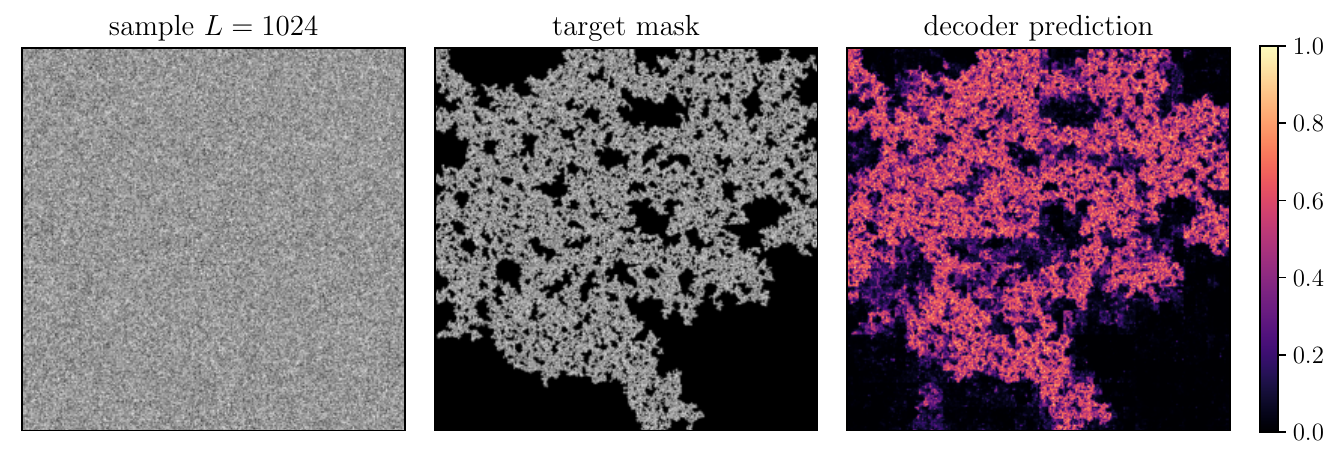}
\caption{
Site-percolation configuration at criticality \textbf{(left)}, mask of the spanning cluster \textbf{(center)}, and prediction of the neural network \textbf{(right)}. The lattice size is $L=1024$, which lies outside the training range and therefore probes finite-size extrapolation.
}
\label{fig:percolation_samples}
\end{figure*}

Percolation is  a central  problem in statistical physics~\cite{stauffer1994introduction,bunde1991percolation,grimmett1999percolation}, and offers a minimal and well-controlled setting in which to probe this issue~\cite{bunde2012fractals,cardy1996scaling}. In two-dimensional site percolation, the existence of a spanning cluster is a global connectivity property, while critical configurations are fractal and governed by universal scaling laws. Recent ML studies have shown that standard convolutional neural networks can roughly identify the percolation threshold or infer the occupation density, but they are not able to recognize the percolating cluster itself~\cite{zhang2019machine,shen2021supervised,cheng2021machine,bayo2023percolating,bayo2025machine}. This suggests that the standard networks may learn density-like local proxies rather than the scale-dependent connectivity structure that defines the physical transition.

This observation connects naturally with the renormalization group (RG), whose purpose is precisely to describe how relevant degrees of freedom evolve under changes of scale~\cite{barenblatt2003scaling,wilson1974renormalization,goldenfeld2018lectures}. Several works have explored the relation between deep learning and RG, including mappings between restricted Boltzmann machines and variational RG~\cite{mehta2014exact}, information-theoretic neural RG schemes~\cite{koch2018mutual}, normalizing-flow-based RG approaches~\cite{li2018neural}, neural ODE for functional RG approaches~\cite{di2022deep} when the Hamiltonian is given and more recently, generative perspectives~\cite{hou2023machine,masuki2025generative,sheshmani2025renormalization,marchand2023multiscale}. However, much of this literature focuses on unsupervised learning or on generative coarse-graining, often using the two-dimensional Ising model as a testbed~\cite{kaur2026can}. Capturing the correct critical exponents and anomalous scaling behavior seems to remain out of reach for present approaches. 
From this perspective, RG provides more than an analogy with deep learning. It suggests a concrete architectural principle: the same local transformation should be applied recursively across scales, so that the representation evolves through a hierarchy of coarse-grained variables. For critical percolation, such a representation should preserve the information relevant to spanning connectivity while discarding microscopic details that are irrelevant under coarse graining. This is precisely the type of inductive bias that is absent from standard convolutional architectures, where filters are local but the learned representation is not constrained to transform consistently across scales.

In this Letter, we investigate this idea in a supervised setting. We consider two related tasks for two-dimensional site percolation: predicting whether a configuration percolates and reconstructing the corresponding spanning cluster. 
To this purpose, we introduce an original scale-shared encoder-decoder architecture in which the same learned coarse-graining (respectively fine-graining) map is applied at every scale. Our results, going well beyond previously considered benchmarks, 
show that the resulting model learns a finite-size scaling order parameter, extrapolates to lattice sizes much larger than those used during training, and reconstructs the nonlocal cluster structure responsible for percolation. In addition, the critical exponents of the theory are faithfully reproduced by the model predictions and remarkably, these performances  seem to rely crucially on the presence of an accurate RG flow in the latent space.

\paragraph*{Percolation model.}
We consider two-dimensional site percolation on a square lattice of linear size $L$. Each site is independently occupied with probability $p$ (also called permeability) and empty otherwise. Occupied nearest-neighbor sites belong to the same cluster. A configuration is said to percolate if at least one occupied cluster spans the system either horizontally or vertically. A representative critical configuration, together with the corresponding spanning-cluster mask and the prediction of our model, is shown in Fig.~\ref{fig:percolation_samples}.
Theoretical results have been obtained in the thermodynamic limit, analytically or numerically~\cite{cardy1992critical,newman2000efficient}; site percolation undergoes a continuous geometric phase transition at the critical occupation probability $p_c \simeq 0.592746$ ,
for the square lattice. Denoting by $\Pi(p,L)$ the probability that a system of size $L$ percolates, one has
$\lim_{L\to\infty} \Pi(p,L)=\Theta(p-p_c).$
Near criticality, the correlation length diverges as
\begin{equation}\label{eq:nu}
\xi(p) \sim |p-p_c|^{-\nu},
\qquad \nu=\frac{4}{3}. 
\end{equation}
The order parameter $P_\infty$, defined as the probability that a site belongs to the infinite cluster, scales for $p>p_c$ as
\begin{equation}\label{eq:beta}
P_\infty(p) \sim (p-p_c)^\beta,
\qquad \beta=\frac{5}{36}.
\end{equation}
For finite systems, the transition is rounded and the relevant scaling variable is $(p-p_c)L^{1/\nu}$. Thus,
$\Pi(p,L) \simeq \mathcal{F}\left((p-p_c)L^{1/\nu}\right)$
while the mass of the critical spanning cluster scales as
$M(L) \sim L^{d_f}$, with fractal dimension $d_f = 2-\frac{\beta}{\nu}=\frac{91}{48}$.
These finite-size scaling laws make percolation a natural benchmark for testing whether a neural network has learned a scale-dependent connectivity observable rather than a local density proxy.

\paragraph*{Methods.}
\begin{figure}[ht]
\includegraphics[width=0.9\columnwidth]{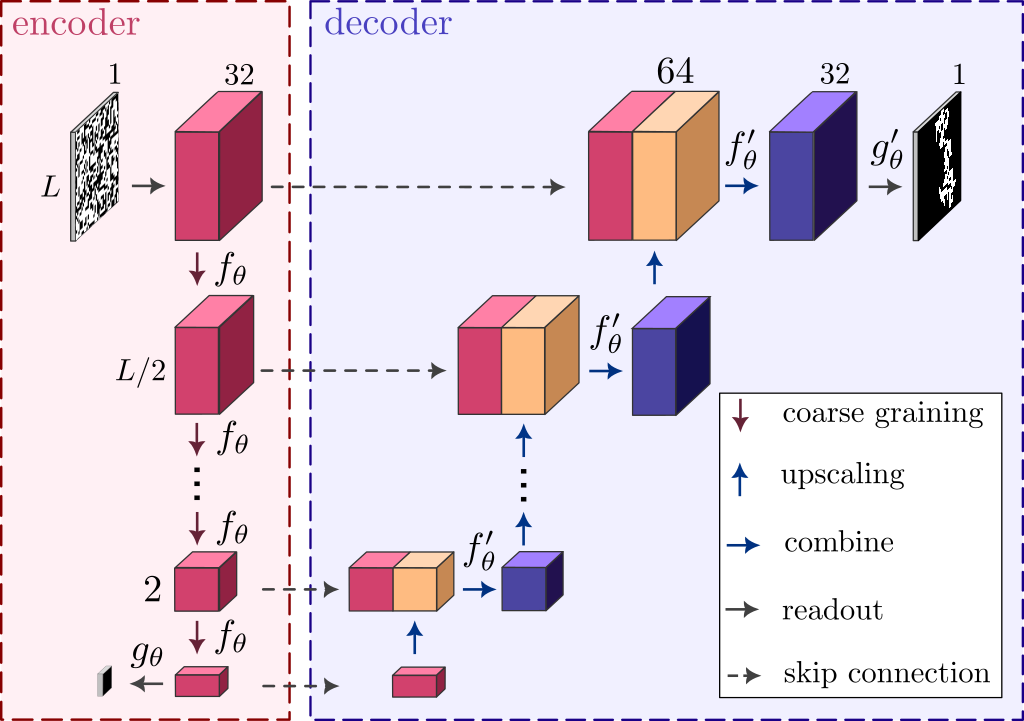}
\caption{
Scale-shared RG-inspired encoder-decoder architecture. The encoder applies the same learned coarse-graining map $f_\theta$ at every scale until the lattice is represented by a single latent vector, from which a readout $g_\theta$ predicts the percolation probability. The decoder applies the same learned fine-graining map $f'_\theta$ at every scale, combining coarse information with encoder features to reconstruct the largest-cluster mask.
}
\label{fig:model_architecture}
\end{figure}
\begin{figure}[t]
\centering
\includegraphics[width=\columnwidth]{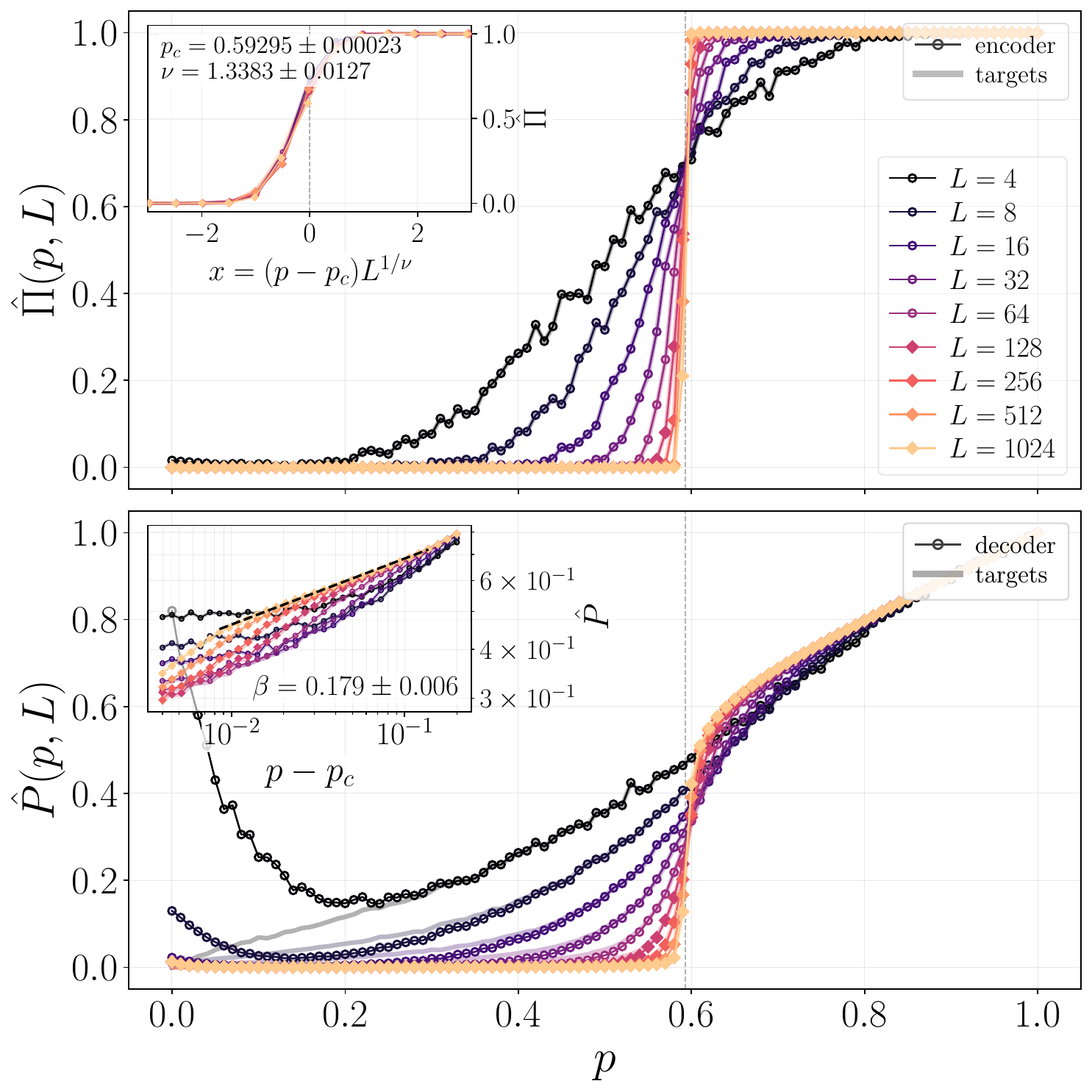}
\caption{
Finite-size scaling of the learned percolation observables. Top Panel: predicted crossing probability $\widehat{\Pi}(p,L)$ as a function of $p$, with inset showing the collapse as a function of $x=(p-p_c)L^{1/\nu}$. Bottom Panel: predicted order parameter $\hat P_\infty(p,L)$ obtained from the decoder output, compared with the target cluster mass fraction. The inset shows the behavior above the critical point and the corresponding estimate of $\beta$. 
}
\label{fig:fss_collapse}
\end{figure}
We introduce a scale-shared encoder-decoder architecture designed to mimic the hierarchical structure of a real-space renormalization procedure. The central constraint is that the same learned coarse-graining rule is applied at every scale. As a consequence, the number of parameters is independent of the lattice size $L$, while larger systems are processed by applying the same transformation more times. This is a variant of a U-Net~\cite{ronneberger2015u} that we specifically adapt to  the RG setting. The architecture is sketched in Fig. \ref{fig:model_architecture}.

Given a percolation configuration $x\in \{0,1\}^{L\times L}$, with $L=2^k$, the encoder first lifts the binary field to a $C$-channel latent representation $h^0$. It then constructs a hierarchy of coarse-grained fields
\begin{equation}\label{eq:coarse-graining}
h^{\ell+1}=f_\theta(h^\ell),
\qquad \ell=0,\ldots,k-1,
\end{equation}
where $h^\ell\in\R^{\left(\frac{L}{2^\ell}\right)^2\times C}$ has spatial size $L/2^\ell$ (pictured as red blocks in the sketch). The map $f_\theta$ is local and reduces the linear resolution by a factor two at each step, while keeping the number of channels fixed. After $k$ iterations, the whole configuration is represented by a single vector $h^k\in\mathbb R^C$. A readout map $g_\theta$ then predicts the probability that the sample percolates,
\begin{equation}
\Pi_\theta = \sigma\left(g_\theta(h^k)\right),
\end{equation}
where $\sigma$ is the sigmoid function, giving for percolation a scalar neural order parameter~\cite{carrasquilla2017machine}.
The encoder is trained using a binary cross-entropy loss on the percolation label.

The decoder is used to reconstruct the largest-cluster mask, which is non empty even for non-spanning configurations.
It starts from the coarsest representation   to proceed toward reproducing the fine-grained details. Denoting the decoder variables by $z^\ell$ (blue blocks in the sketch), we set $z^k=h^k$ and iterate
\begin{equation}\label{eq:fine-graining}
z^\ell
=
{f'}_\theta\left(h^\ell,\mathcal U z^{\ell+1}\right),
\qquad
\ell=k-1,\ldots,0,
\end{equation}
where $\mathcal U$ denotes a fixed upsampling operator. At each scale, the decoder combines the information propagated from coarser scales with the encoder representation at the same resolution. The same fine-graining rule $f'_\theta$ is used at every scale. A final 
 readout predicts the mask
\begin{equation}\label{eq:hpi}
p_{\theta,i}
=
\sigma\left(g'_\theta(z^0_i)\right),
\end{equation}
where $p_{\theta,i}$ is the predicted probability that site $i$ belongs to the largest cluster. The decoder is trained with a pixel-wise binary cross-entropy loss on the true mask.

This construction differs from a standard convolutional encoder-decoder in its explicit scale sharing: the same coarse- and fine-graining maps are reused across all levels of the hierarchy. This makes the architecture naturally applicable to lattice sizes larger than those seen during training. 
The proposed architecture is original and is further described in the End Matter. 

\paragraph*{Results.}
We train the model on a mixed-size ensemble of square lattices with $L=2^k$, $k=2,\ldots,6$, using $2\times 10^4$ samples per size. The trained model is then evaluated on $k=2,\ldots,10$, corresponding to $L=1024$, i.e.~a linear size $16$ times larger than the largest training lattice. Samples are drawn with values of $p$ near criticality, using the finite-size scaling variable
\begin{equation}
\label{eq: FSS p-window}
x = (p-p_c)L^{1/\nu},
\end{equation}
with $p_c=0.592746$ and $\nu=4/3$. For each size, the range of values of $p$ sampled is chosen to cover the rounded transition region, where the crossing probability satisfies approximately $0.1<\Pi(p,L)<0.9$. This avoids trivial examples far from criticality.
This combination of sizes and $p$'s makes both the classification and reconstruction tasks sensitive to long-range connectivity.
We first examine whether the network output has the expected finite-size scaling structure. Let $\widehat{\Pi}(p,L)$ denote the empirical average of the predicted probability of percolation for samples of size $L$ and occupation probability $p$. As shown in Fig.~\ref{fig:fss_collapse}, $\widehat{\Pi}(p,L)$ develops a sharp transition near the known critical point. When plotted as a function of $x=(p-p_c)L^{1/\nu}$, the curves obtained for different lattice sizes collapse onto a common scaling function, including sizes not seen during training. A finite-size scaling fit  $p_c^{\mathrm{fit}}\simeq 0.59295$ and $\nu^{\mathrm{fit}}\simeq 1.3383$, 
quite close to the two-dimensional percolation values.
This indicates that the network has learned a dimensionless crossing/no-crossing observable, rather than a size-dependent decision boundary.

The decoder output gives access to a spatially resolved estimate of the largest-cluster mask. For $p>p_c$, the predicted cluster mass fraction
$\hat P_\infty(p,L)
=
{1}/{L^2}
\sum_i \hat p_i$
provides a finite-size estimate of the percolation order parameter. As shown in Fig.~\ref{fig:fss_collapse}, this quantity follows the expected growth above the transition. A fit of $\hat P_\infty(p,L)$ above $p_c$ at the largest tested size gives an effective exponent $\beta^{\rm fit}\simeq 0.179\pm 0.006$, while the scaling of $\hat P_\infty(p_c,L)$ gives a fractal dimension $d_f\simeq 1.887$, equivalently $\beta/\nu\simeq 2-d_f$. These estimates are close to the exact values $\beta= 5/36 \approx 0.139$ and $d_f =91/48 \approx 1.896$, within the finite-size and learning errors of the experiment.

\begin{figure}[h]
\centering
\includegraphics[width=\columnwidth]{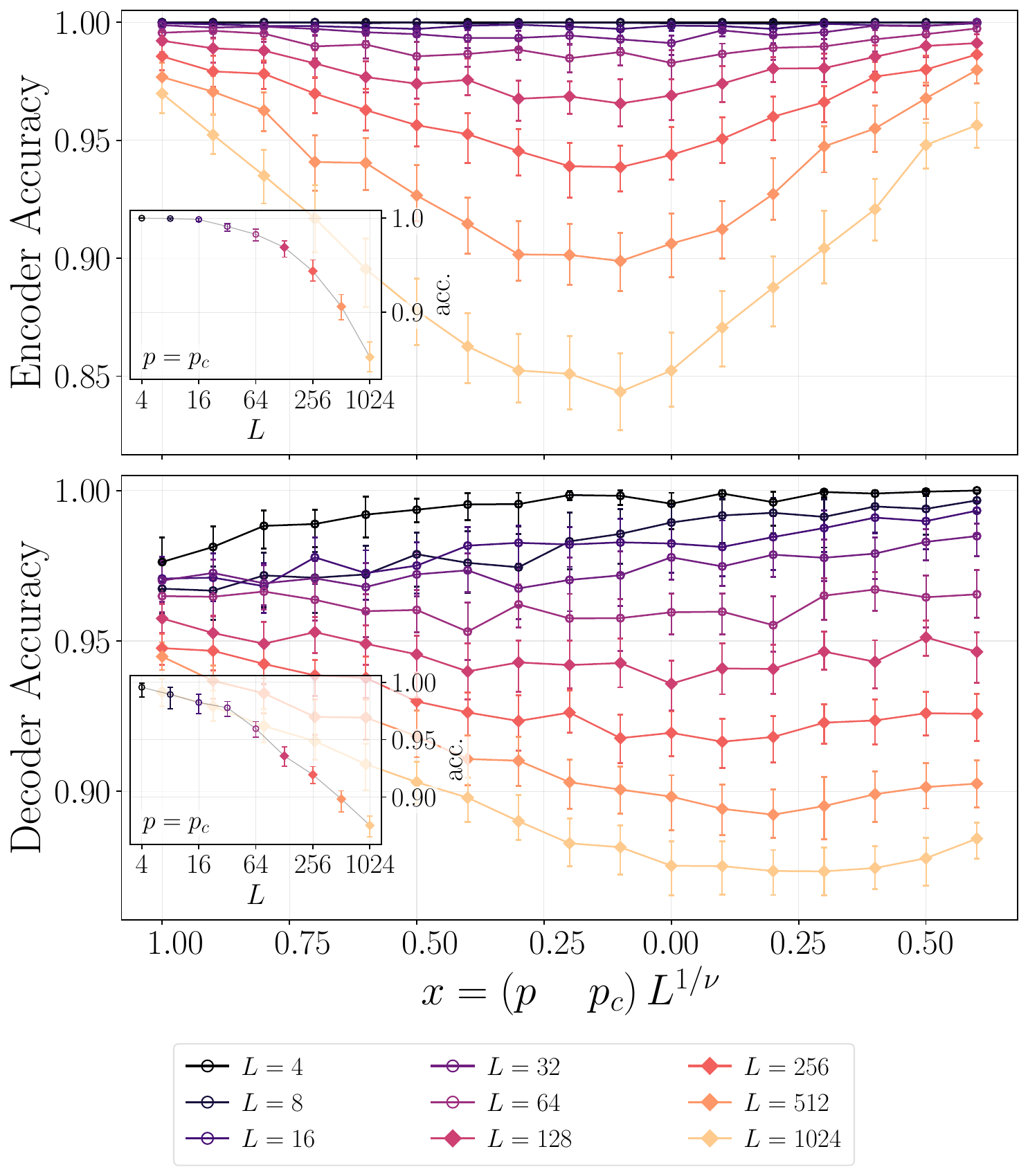}
\caption{
Classification and reconstruction performance as a function of the rescaled occupation probability $x=(p-p_c)L^{1/\nu}$. Top: encoder accuracy for the percolation label. Bottom: decoder accuracy for the largest-cluster mask. The largest size used during training is $L=64$, while the test sizes extend up to $L=1024$. For each value of $x$, the reported value is the median over $N=5000$ test samples. Error bars indicate bootstrap uncertainty over test samples. Insets show the performance at criticality as a function of $L$.
}
\label{fig:encoder_accuracy}
\end{figure}
To corroborate quantitatively these findings, we then evaluate the supervised tasks directly. Figure~\ref{fig:encoder_accuracy} reports the classification accuracy of the encoder and the reconstruction accuracy of the decoder across system sizes and values of $x$. The performances of this architecture are quite striking. The model maintains high classification accuracy across the critical window, including in the extrapolation regime $L>64$. As expected, the accuracy is lowest near $p_c$, where microscopic changes can alter the global connectivity label. 
By contrast, as shown in SM a standard U-Net achieves comparable performance on lattice
sizes represented during training, but it deteriorates sharply
when extrapolating to larger systems. Thus, successful interpolation of the
percolation transition does not by itself guarantee scale generalization;
this comparison supports scale sharing as the key inductive bias enabling
extrapolation across system sizes.

For cluster reconstruction, the encoder-decoder is trained directly from scratch using a pixel-wise binary cross-entropy loss against the largest-cluster mask. This strategy outperforms freezing a classifier-pretrained encoder, likely because the classification task emphasizes the existence of a spanning cluster, whereas the reconstruction task requires identifying the largest cluster in both percolating and non-percolating configurations. 
The reconstruction task is more stringent than binary classification: a classifier may exploit finite-size texture gradients, whereas reconstructing the cluster requires identifying a nonlocal object across scales. The example shown in Fig.~\ref{fig:percolation_samples} confirms that the decoder preserves the coarse connectivity of the spanning structure and refines its geometry down to the microscopic scale, even for $L=1024$.

Together, these observations show that the scale-shared architecture learns more than an occupation-density proxy. Its global output obeys finite-size scaling, its decoder recovers the spatial structure of the largest cluster, and both properties persist beyond the training range.

\paragraph*{Latent RG flow.}
\begin{figure}[h]
  \centering
  \includegraphics[width=0.9\columnwidth]{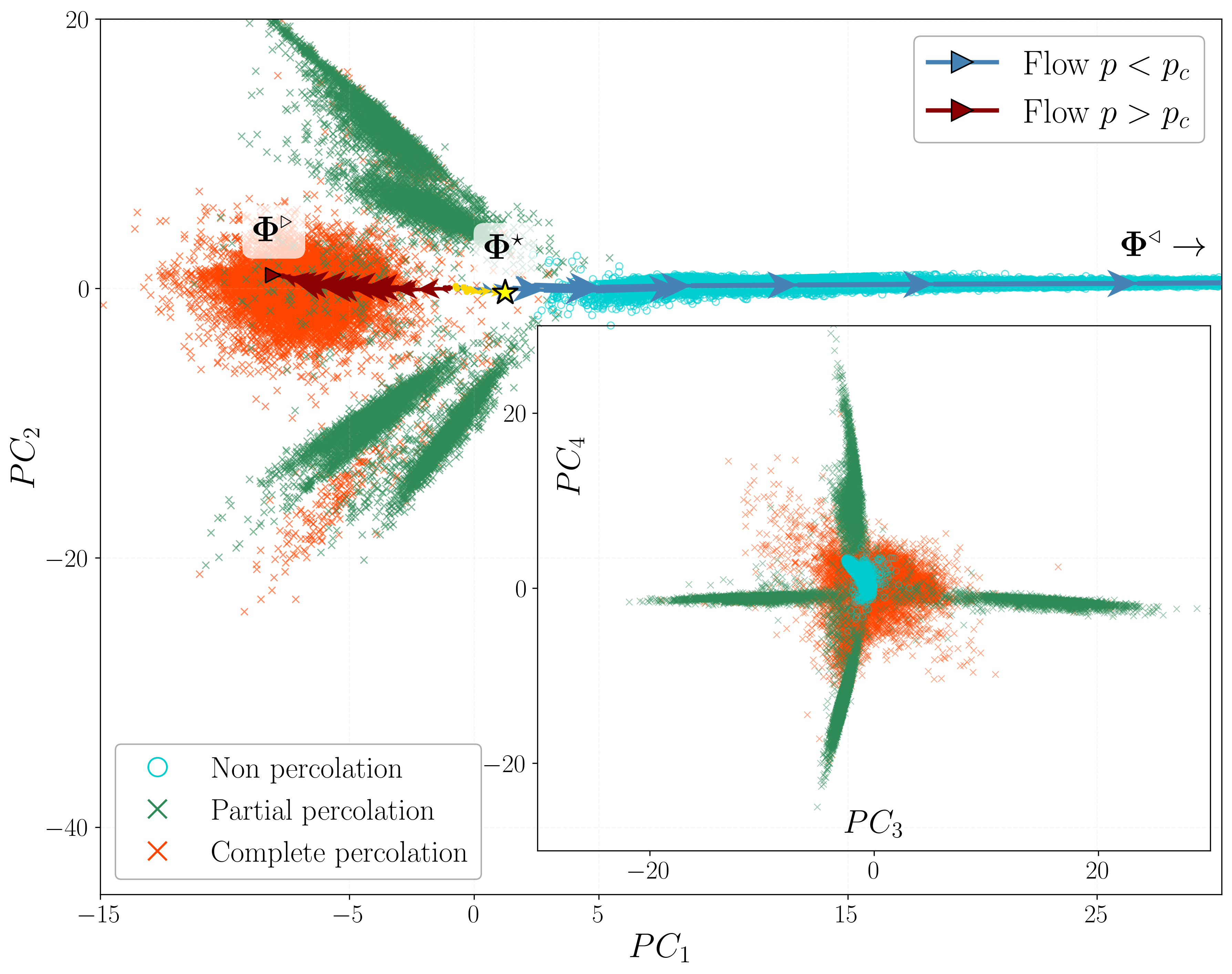}
  \vspace{0.5cm}
\includegraphics[width=0.9\columnwidth]{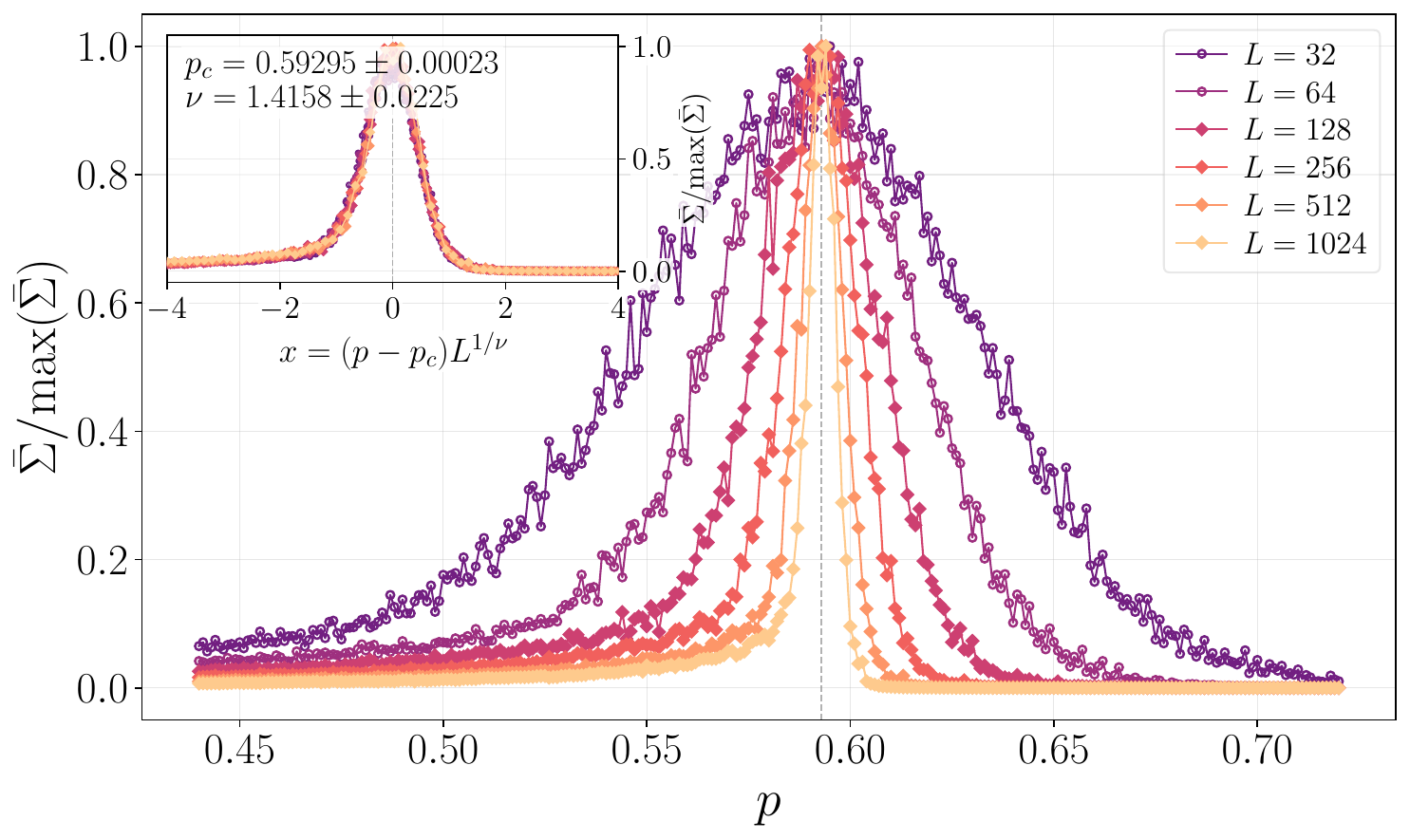}
\caption{
Top:Projection of the coarsest latent representation $\Phi$ onto principal components computed from critical samples at the largest training size. Points show critical samples. Colors correspond to distinct connectivity patterns, including non-percolating configurations (light blue), configurations spanning in both directions (orange), and other percolating patterns related by rotations (green). 
Six clusters are clearly apparent.
Arrows indicate the motion of the distribution $\Phi(p,L)$'s centroid as the lattice size increases, for values of $p$ below and above $p_c$, providing an empirical visualization of the learned scale flow. Bottom: Finite-size scaling of the normalized latent susceptibility $\bar\Sigma(p,L)$ as function of $p$ with its collapse inset as function of $x=(p-pc)L^{1/\nu}$ 
}
\label{fig:RG-flow}
\end{figure}
We finally analyze how the scaling behavior is represented internally by the encoder. Let
\begin{equation}
\Phi = h^k \in \mathbb R^C
\end{equation}
be the coarsest latent vector obtained after $k=\log_2 L$ applications of the same learned coarse-graining map. Since increasing $L$ amounts to applying the same map more times, the family of distributions
$\rho_{\theta,L}(\Phi\mid p)$
provides an empirical finite-dimensional representation of the learned flow with scale.
The covariance matrix $\Sigma(\Phi \mid p,L)$ of this field already indicates that its distribution  carries relevant information about it. 
For fixed $p$ and $L$, we estimate the covariance of $\Phi$ and define the 
latent susceptibility
\begin{equation}
\overline{\Sigma}_\theta(p,L) = \operatorname{Tr}
\left[
\Sigma(\Phi \mid p,L)
\right],
\end{equation}
which measures sample-to-sample fluctuations of the coarse-grained representation. As shown in Fig.~\ref{fig:RG-flow}, $\Sigma_\theta(p,L)$ peaks near the transition. Its normalized profiles collapse under finite-size rescaling, with fitted values $p_c^{\mathrm{fit}}\simeq 0.59295$ and $\nu^{\mathrm{fit}}\simeq 1.4158$.
Thus, the latent representation itself carries critical fluctuations with the expected scaling structure.
This provides an independent confirmation that the network has learned the whole structure of the transition.

To visualize the learned flow, we project $\Phi$ onto the leading principal components computed from critical samples at the largest training size. Figure~\ref{fig:RG-flow} shows that the critical latent distribution is a mixture of six  components associated with distinct connectivity patterns: one for non-percolating configurations, one for percolating configurations in both directions, and four components for percolating configurations in two or three directions and all the rotations of such patterns. This organization indicates that the network has learned a representation sensitive not only to the existence of percolation, but also to the topology of boundary connections.

Changing $p$ away from $p_c$ deforms this latent distribution. As $L$ increases, the centroid of the projected distribution  $\Phi(p,L)$ drifts for $p>p_c$ [resp. $p<p_c$], toward the percolating [resp. non-percolating] sector materialized by the stable fixed point $\Phi^{\rhd}$ [resp. $\Phi^{\lhd}$]. The critical distribution lies near the separatrix between these two trends, with an unstable centroid fixed point $\Phi^\star$. This behavior is consistent with the RG picture of an unstable critical fixed point separating two stable phases, although the present analysis should be understood as a finite-dimensional learned proxy rather than a direct reconstruction of the full RG transformation.

\paragraph*{Conclusion.}
We have shown that a scale-shared encoder-decoder architecture can learn critical percolation observables in a supervised setting and extrapolate to lattice sizes larger than those used during training. The model does not only classify percolation: it reconstructs the spatial cluster structure and produces observables whose finite-size scaling are in agreement with that of two-dimensional percolation. 
Our method goes beyond phases classification with a black-box scalar neural order parameter~\cite{carrasquilla2017machine}, by learning a neural order parameter in the form of a latent field following the RG structure of a critical system.  This further provides an empirical visualization of a learned scale flow between non-percolating and percolating sectors.
These results point out that RG-inspired weight sharing is a key inductive bias for learning scale-invariant systems, especially when the relevant observables are nonlocal. A more complete reconstruction of the learned RG dynamics, including symmetry constraints and a controlled treatment of latent-field normalization, remains an important direction for future work.

\paragraph*{Acknowledgments}
    Authors wish to thank Raoul Santachiara and Francesco Chippari  for insightful discussions on percolation;  We also warmly thank Beatriz Seoane and Aurelien Decelle for fruitful discussions. We acknowledge financial support by the French ANR grant Scalp (ANR-24-CE23-1320).

\bibliographystyle{unsrt}
\bibliography{prl}


\onecolumngrid

\section*{End Matter}

\paragraph*{Architectural details.}
We describe here the local parametrization of the scale-shared encoder--decoder used in the experiments. The same encoder rule $f_\theta$ and decoder rule $f'_\theta$ are applied at every scale. Consequently, the number of trainable parameters does not depend on the lattice size $L$.

\paragraph*{Encoder.}
The coarse-graining map in Eq.~\eqref{eq:coarse-graining} is written as
\begin{equation}
f_\theta(h^\ell)
=
\phi_\theta\left(
\mathrm{ReLU} \circ
\operatorname{Conv}_\theta(h^\ell)
\right),
\end{equation}
where ReLU denotes the Rectified Linear Unit nonlinearity, applied element-wise.
The first operation is a standard two-dimensional convolution with kernel size $3\times 3$, stride $1$, mirror padding at boundaries, and $C$ input and output channels. Its role is to let neighboring blocks exchange information before the actual coarse-graining step. This operation is important in practice, since treating $2\times 2$ blocks independently would lose information at block boundaries.
The second operation, $\phi_\theta$, performs the block collapse. 
The field $\operatorname{Conv}_\theta(h^\ell)$ is partitioned into non-overlapping $2\times 2$ patches. On each patch, the $4C$ input values are mapped to one $C$-dimensional vector by the same two-layer multilayer perceptron (MLP),
\begin{equation}
\phi_\theta:\mathbb R^{4C}\to\mathbb R^C.
\end{equation}
In all experiments we use $C=32$ channels, ReLU activations and hidden width $64$.
Since the $2\times 2$ patches are reduced to a single site, each application of $f_\theta$ divides the number of sites by $4$ and halves the linear system size. Iterating this rule $k=\log_2 L$ times maps the original configuration to a single latent vector $h^k\in\mathbb R^C$. 
The final encoder readout $g_\theta$ is a small multilayer perceptron (one hidden layer of hidden width $64$):
$
\Pi_\theta
=
\sigma\left(g_\theta(h^k)\right)
$, where $\Pi_\theta$ is interpreted as the predicted probability that the configuration percolates. 
The encoder is trained with a binary cross-entropy loss on the percolation label.

\paragraph*{Decoder.}
The decoder starts from the coarsest representation $z^k=h^k$ and refines it scale by scale. At each level, the coarse decoder representation is first upsampled by copying each coarse pixel onto the corresponding $2\times 2$ block,
\begin{equation}
\mathcal U z^{\ell+1}
\in
\mathbb R^{C\times (L/2^\ell)\times (L/2^\ell)} .
\end{equation}
It is then concatenated with the encoder feature $h^\ell$ at the same resolution. The fine-graining rule
$
z^\ell
=
f'_\theta\left(
h^\ell,\mathcal U z^{\ell+1}
\right)
$ (Eq.~\ref{eq:fine-graining})
is applied independently on each $2\times 2$ patch. It applies a linear map to the $8C$ values contained in the concatenated patch to output $4C$ values, which are reshaped into a refined $C$-channel $2\times 2$ block, followed by a ReLU nonlinearity.
Equivalently, $f'_\theta$ may be viewed as a local $2\times 2$ (stride $2$) convolution followed by a nonlinearity.
The skip connection through $h^\ell$ plays the same role as in a U-Net: it reintroduces information recorded by the encoder at resolution $\ell$, which cannot be recovered from the coarser representation alone. The difference with a standard U-Net is that the same rule $f'_\theta$ is shared across all scales.
At the final scale, the decoder output is combined with the original binary configuration $x$ and mapped linearly to a four-pixel output on each $2\times 2$ block. This gives the mask prediction~(\ref{eq:hpi})
where $p_{\theta,i}$ can be interpreted as the predicted probability that site $i$ belongs to the largest cluster. The decoder is trained with a pixel-wise binary cross-entropy loss against the largest-cluster mask.

\paragraph*{Parameter count.}
For $C=32$ and hidden width $64$, the scale-shared encoder rule $f_\theta$ contains $19{,}584$ parameters, while the classifier readout $g_\theta$ contains $2{,}177$ parameters. The decoder fine-graining rule $f'_\theta$ contains $32{,}896$ parameters and the final decoder readout  $g'_\theta$  contains $532$ parameters. The full encoder--decoder therefore contains $55{,}189$ trainable parameters. Since all local rules are shared across scales, this number is the same for all lattice sizes.


\end{document}